\documentclass[
onecolumn]{IEEEtran}
\ifCLASSINFOpdf
\else
\fi
\usepackage{settings}

\begin{document}

\title{
A Multi-Scale Spatiotemporal Perspective of Connected and Automated Vehicles: Applications and Wireless Networking
}
%
%
%
%

\author{
\IEEEauthorblockN{Rajit Johri$^\dag$, 
Jayanthi Rao$^\dag$, Hai Yu$^\dag$, Hongwei Zhang$^*$\thanks{The majority of this work was done when Hongwei Zhang visited Ford Motor Company on his sabbatical. Authors are in alphabetic order, with Hongwei Zhang (hongwei@wayne.edu) being the correspondence author. Zhang's work is supported in part by Ford Research and NSF awards CNS-1136007, CNS-1054634, GENI-1890, and GENI-1633.} \\ }
\IEEEauthorblockA{
$^\dag$Research and Advanced Engineering, Ford Motor Company \\ 
$^*$Department of Computer Science, Wayne State University \\
\{rjohri,
jrao1,hyu20\}@ford.com, hongwei@wayne.edu \\
}

}

\maketitle

\begin{abstract} 
Wireless communication is a basis of the vision of connected and automated vehicles (CAVs). Given the heterogeneity of both wireless communication technologies and CAV applications, one question that is critical to technology road-mapping and policy making is which communication technology is more suitable for a specific CAV application. 
	Focusing on the technical aspect of this question, we present a multi-scale spatiotemporal perspective of wireless communication technologies as well as canonical CAV applications in active safety, fuel economy and emission control, vehicle automation, and vehicular infotainment. 
	Our analysis shows that CAV applications in the regime of small spatiotemporal scale communication requirements are best supported by V2V communications, applications in the regime of large spatiotemporal scale communication requirements are better supported by cellular communications, and applications in the regime of small spatial scale but medium-to-large temporal scale can be supported by both V2V and cellular communications and provide the opportunity of leveraging heterogeneous communication resources.
\end{abstract}

\section{Introduction}  \label{sec:intro}

Transforming the traditional paradigm of single-vehicle-oriented optimization and operation, next-generation vehicles are expected to coordinate with one another, transportation infrastructures, Internet clouds, and people in maximizing the safety, sustainability, and comfort of road transportation. One basic enabler of this vision of connected and automated vehicle (CAV) operation is for vehicles to wirelessly communicate with one another, transportation infrastructures, Internet clouds, and people. 
	Given the wide spectrum of available wireless communication technologies (e.g., cellular and DSRC) 
and the heterogeneity of CAV applications envisioned, there has been heated debate on the exact wireless communication technologies to be used for CAVs, and complex factors such as market penetration of new technologies have complicated this debate. 
	Towards a thorough technical examination of the debate and for enabling strategic planning of technology roadmap, we analyze different categories of CAV applications by examining their spatiotemporal scales of communication requirements, and we analyze capacity limits of different wireless communication alternatives. Based on this analysis, we make recommendations on technology planning. 

The rest of this article is organized as follows. 
	In Section~\ref{sec:apps}, we review representative CAV applications and their communication requirements. In Section~\ref{sec:networks}, we comparatively analyze cellular communication and V2V communication in supporting CAV applications. We make concluding remarks in Section~\ref{sec:concludingRemarks}.

\section{Connected and automated vehicle applications}  \label{sec:apps}

Through wireless communication with one another, transportation infrastructures, Internet clouds, and people, connected and automated vehicles (CAVs) will optimize their operations for safety, sustainability, and user comfort by treating themselves as integral parts of the transportation ecosystem, and they will also serve as enablers and active participants in new paradigms of road transportation. 
	For instance, wireless communication enables non-line-of-sight sensing among vehicles and transportation infrastructures, thus effectively extending the sensing range of what traditional radars and cameras can enable. 
	Wireless communication enables vehicles to share in real-time their internal operation state (e.g., hard brake) that were infeasible before \cite{Wang:platoonDelay}. 
	Wireless communication enables vehicles to exchange their knowledge about traffic condition, road condition, and the environment, thus enabling real-time, extended sensing of driving conditions as well as data-driven transportation planning. 
	Wireless communication also connects vehicles and people in a seamless manner, thus enabling transformative user experience. 

In what follows, we review representative CAV applications in the broad areas of active safety, fuel economy and emission control, automation, and infotainment. When presenting the type of communication required by a CAV application (e.g., in Tables~\ref{table:safety}-\ref{table:infotainment}), we use ``V'', ``I'', ``C'', and ``M'' to denote ``Vehicle'', ``Infrastructure'', ``Internet cloud'', and ``Mobile device'' respectively.


\subsection{Active safety} \label{subsec:activeSafety}

Enabling vehicles and transportation infrastructures to sense and coordinate with one another, wireless communication can help prevent up to 80\% of today's accidents \cite{activeSafety-req}. In particular, eight high-priority active safety applications identified by NHTSA and VSC are as follows 
\ifthenelse{\boolean{short}}
{\cite{activeSafety-req}:}
{\cite{activeSafety-req,NHTSA2005-itsReport}:}
\begin{itemize}
\item \emph{Pre-crash sensing (PCS)}: it is used to prepare for imminent, unavoidable collisions.

\item \emph{Emergency brake lights (EBL)}: When a vehicle brakes hard, it sends a message to other vehicles following behind.

\item \emph{Collaborative collision warning (CCW)}: it collects surrounding vehicle locations and dynamics and warns the driver when a collision is likely. 

\item \emph{Left turn assistant (LTA)}: it provides information to drivers about oncoming traffic to help them make a left turn at a signalized intersection without a phasing left turn arrow.

\item \emph{Lane change warning (LCW)}: it provides a warning to the driver if an intended lane change may cause a crash with a nearby vehicle.

\item \emph{Traffic signal violation warning (TSV)}: it uses infrastructure-to-vehicle communication to warn the driver to stop at the legally prescribed location if the traffic signal indicates a stop and it is predicted that the driver will be in violation. 

\item \emph{Stop sign violation warning (SSV)}: it uses infrastructure-to-vehicle communication to warn the driver if the distance to the legally prescribed stopping location and the speed of the vehicle indicate that a relatively high level of braking is required for a complete stop.

\item \emph{Curve speed warning (CSW}: it aids the driver in negotiating curves at appropriate speeds. 

\end{itemize}
Table~\ref{table:safety} summarizes the communication requirements for these active safety applications.
\begin{table}[!htbp]
\begin{center}
\begin{tabular}{| l | c | c | c | c |}
\hline
Application  &  Comm. type  &   Range   &   Frequency   &  Tolerable latency \\ 
\hline
PCS  &  V2V   &  50m  &  50Hz    &  20msec  \\
\hline
EBL  &  V2V   &  200m  &  10Hz    &  100msec   \\
\hline
CCW  &  V2V   &  150m  &  10Hz    &  100msec   \\
\hline
LTA  &  I2V, V2I   &  300m  &  10Hz    &  100msec   \\
\hline
LCW  &  V2V   &  150m  &  10Hz    &  100msec   \\
\hline
TSV  &  I2V   &  250m  &  10Hz    &  100msec   \\
\hline
SSV  &  I2V, V2I   &  300m  &  10Hz    &  100msec   \\
\hline
CSW  &  I2V   &  200m  &  1Hz    &  1000msec   \\
\hline
\end{tabular}
\end{center}
\caption{Communication requirements of active safety applications \cite{activeSafety-req}}  \label{table:safety}
\end{table}

\subsection{Fuel economy and emission control} \label{subsec:fuel}

By enabling sensing of traffic, road, infrastructure, and environmental conditions at different spatiotemporal scales, wireless communication enables the paradigm shift from single-vehicle-centered fuel economy and emission control to networked control. A few canonical examples are as follows:
\begin{itemize}
\item \emph{Low emission zone (LEZ)}: it lets a hybrid electrical vehicle (HEV) 
 operate in the minimal-greenhouse-gas-emission mode in certain geographical zones (i.e., low emission zones) specified by GPS data \cite{green-zone}. The low emission zones can be specified by governments, organizations, and/or individuals. 

\item \emph{Fuel-economy-oriented trip planning (FEP)}: it performs trip-wide, global energy usage optimization based on trip previewing and profiling. It includes both mode control and continuous control (e.g., torque control, torque allocation to different actuators such as the mechanical and the electrical motors), and the control considers optimization over a spatial scale ranging from hundreds of feet to tens of miles 
\ifthenelse{\boolean{short}}
{\cite{Ekici:VNET-survey}.}
{\cite{Zhao:route-guidance,Ekici:VNET-survey}.}

\item \emph{Preview-based fuel economy optimization (PFO)}: it optimizes hybrid electrical vehicle (HEV) powertrain based on previewed road information, and it focuses on instantaneous mode and continuous control 
\ifthenelse{\boolean{short}}
{\cite{Johri:fuel-emission,Peng:PHEV-powerManagement}. }
{\cite{Johri:fuel-emission,Peng:PHEV-powerManagement,Kolmanovsky:HEV-gameControl,Onder:powerSplitControl,Peng:truck-powerManagement,Staccia:HEV-powerManagement}. } 
(Note: unlike PFO, FEP is for PHEV with additional/more battery than HEV.)

\item \emph{Stop-start}: to save fuel, it stops/starts traditional engines based on surrounding vehicle state and immediate traffic regulation state, for instance, temporarily stopping an engine in front of red light 
\ifthenelse{\boolean{short}}
{\cite{Kuang:start-stop}. }
{\cite{Kuang:start-stop,Semar:start-stop,Kees:micro-hybrid}. }

\item \emph{Platoon}: it organizes vehicles into platoons to reduce the air drag for the lead and following vehicles so as to save fuel/energy while ensuring safety 
\ifthenelse{\boolean{short}}
{\cite{Ekici:VNET-survey,Wang:platoonPDR}. }
{\cite{Ekici:VNET-survey,Tsuagawa:platoonOverview,Zhang:platoonDemon,Wang:platoonPDR,Wang:platoonDelay}. }

\end{itemize}
Based on typical operating scenarios of these fuel economy and emission control applications, Table~\ref{table:fuel} summarizes their communication requirements.
\begin{table}[!htbp]
\begin{center}
\begin{tabular}{| l | c | c | c | c |}
\hline
Application  &  Comm. type  &   Range   &   Frequency   &  Tolerable latency \\
\hline
LEZ  &  C2V,V2C   &  $\ge$1km  &  $\le$1Hz, e.g., 0.1Hz    &  $\ge$10sec  \\
\hline
FEP   &   C/I/V2V,V2V/I/C   &   $\ge$300m    &    $\ge$1Hz    &     $\sim$1000msec   \\
\hline
PFO   &   C/I/V2V,V2V/I/C   &   300m-13km   &   $\ge$1Hz    &    distance-dependent  \\
\hline
Stop-start  &  V2V,V2I,I2V    &   $\sim$50m    &   $\ge$1Hz     &   $\sim$100msec  \\
\hline
Platoon   &    V2V     &    50m-300m    &    $\ge$10Hz     & $\sim$100msec \\
\hline
\end{tabular}
\end{center}
\caption{Communication requirements of fuel economy and emission control applications}  \label{table:fuel}
\end{table}

\subsection{Networked vehicle automation}  \label{subsec:automation}

Besides networked fuel economy and emission control as discussed in Section~\ref{subsec:fuel}, other advanced vehicle automation features include the following:
\begin{itemize}
\item \emph{Adaptive vehicle tuning (AVT)}: it adjusts the operation state (e.g., refining vehicle suspension and driveline parameters and pre-boosting brake and electronic stability control) of a following vehicle based on real-time information about the operation state of leading vehicles as well as the ensuing road conditions (e.g., speed bumps and potholes) 
\ifthenelse{\boolean{short}}
{\cite{SAFESPOT}. }
{\cite{SAFESPOT,AV:system-algorithms,C2C-CC,CMU-UV,AV:vision}. }

\item \emph{Collaborative failure-mode-effect-management (CFEM)}: In the event of sensor failures on autonomous vehicles, wireless communication enables functioning vehicles to share sensing results with vehicles of mal-functioning sensors 
\ifthenelse{\boolean{short}}
{\cite{SAFESPOT,Ekici:VNET-survey}. }
{\cite{SAFESPOT,AV:system-algorithms,Ekici:VNET-survey}. }

\item \emph{Traffic jam assist (TJA)}: it enables automated driving in traffic jam with stop-and-go and lane keeping 
\ifthenelse{\boolean{short}}
{\cite{UCB-TR-safety}. }
{\cite{UCB-TR-safety,pathPlanning}. }

\item \emph{Supercruise with opportunistic lane change (SCLC)}: it enables automated driving on highways while allowing for lane changes 
\ifthenelse{\boolean{short}}
{\cite{UCB-TR-safety}. }
{\cite{UCB-TR-safety,Arne:microscopidModeling}. }

\item \emph{Automated valet/street parking (AP)}: it enables automated valet or street parking without involving drivers 
\ifthenelse{\boolean{short}}
{\cite{autoParking}. }
{\cite{autoParking,Bosch:fullyAutoParking}. }
\end{itemize}
Based on typical operating scenarios of these vehicle automation applications, Table~\ref{table:automation} summarizes their communication requirements.
\begin{table}[!htbp]
\begin{center}
\begin{tabular}{| l | c | c | c | c |}
\hline
Application  &  Comm. type  &   Range   &   Frequency   &  Tolerable latency \\ 
\hline
AVT   &   V2V,V2C2V    &   $\sim$100m or context (e.g. speed,    &  $\sim$10Hz   & $\sim$100msec    \\
      &               & inter-vehicle gaps) dependent      &    &   \\
\hline
CFEM  &  V2V,V2I,I2V   &  $\sim$300m   &    $\sim$10Hz    &    $\sim$100msec \\
\hline
TJA   &  V2V  &   10m-300m   &    $\sim$10Hz    &     $\sim$100msec \\
\hline
SCLC  &  V2V  &   10m-300m   &    $\sim$10Hz    &     $\sim$100msec \\
\hline
AP    & V2V,V2C2V,V2Sensor  &  10m-1km+     &   $\le$1Hz     &   $\ge$1sec \\
\hline
\end{tabular}
\end{center}
\caption{Communication requirements of vehicle automation applications}  \label{table:automation}
\end{table}

\subsection{Vehicular infotainment}  \label{subsec:infotainment}

By interconnecting vehicles with one another, the transportation infrastructures, the Internet, and people, wireless communication enables networked infotainment services that significantly enhance user experience and vehicle operation. Six canonical examples of user-centric infotainment applications are as follows:
\begin{itemize}
\item \emph{Remote vehicle status (RVS)}: a mobile device (e.g., smartphone, tablet, or smartwatch) application is used to check status of certain vehicle parameters such as its state of charging and location. This is particularly useful for electric vehicles to know the current state of charging, for instance, when parked in a location without being plugged in (e.g., in shopping malls and airports) or plugged in (e.g., at work or other charging facilities) \cite{EV-chargingStation-search}.

\item \emph{Remote vehicle command (RVC)}: a mobile device application is used to send commands to a vehicle to perform functions such as door lock/unlock, remote-start, climate-control, and vehicle charging policy update 
\ifthenelse{\boolean{short}}
{\cite{MyBMWRemoteApp}. }
{\cite{MyBMWRemoteApp,TeslaCommandFromWatch}. }

\item \emph{Real-time cloud services to vehicle (RTCS)}: it enables HMI or powertrain information (e.g., fuel economy) to be served from the cloud in a context-aware manner \cite{cloudVehicleControl}.

\item \emph{Crowd-sourced sensing (CSS)}: it enables Waze-like applications for sensing traffic, air-quality, noise, parking, fuel consumption, camera feeds, etc \cite{USDOT:SmartCity-ConnectedTransportation}.

\item \emph{Delay-insensitive downloads (DID)}: it enables downloading multimedia content (e.g., audio, video, movies) requested by users \cite{Ekici:VNET-survey}.

\item \emph{Geographic proximity applications (GPA)}: it includes, among others, bringing real-time location-specific information (e.g., business ads, coupons, free parking slots) to users and enabling caravaning for group-travelers (e.g., follow-me, group chat) \cite{Jayanthi:VANET-groupApps}.
\end{itemize}
Two canonical examples of vehicle-centric infotainment applications are as follows:
\begin{itemize}
\item \emph{Extended surrounding sensing (ESS)}: it collects surrounding environment and road conditions by integrating the sensing views of multiple vehicles. The extended surrounding sensing information can be used for both driver assistance (e.g., providing 360-degree views) and vehicle control \cite{roadwayActiveSensing}. 

\item \emph{Cloud-based diagnostics/prognostics/analytics (CDPA)}: Data is collected in the cloud regarding certain interesting events or periodic parameters pertaining to vehicle state such as powertrain parameters, fuel economy, diagnostics trouble code, HMI usage statistics, and vehicle dynamic state (e.g., location, heading, velocity, and acceleration) \cite{USDOT:SmartCity-ConnectedTransportation}.

\end{itemize}
Based on typical operating scenarios of these vehicular infotainment applications, Table~\ref{table:infotainment} summarizes their communication requirements.
\begin{table}[!htbp]
\begin{center}
\begin{tabular}{| l | c | c | c | c |}
\hline
Application  &  Comm. type  &   Range   &   Frequency   &  Tolerable latency \\
\hline
RVS   &  V2C,C2V,C2M    & cellular radius,   &  $\le$0.1Hz  &  best-effort \\
      &                 & vehicle-user distance  &   & \\
\hline
RVC   & M2C,C2V,V2C,C2M  &  cellular radius,   &  event-driven  &  $\le$5sec \\
      &                 & vehicle-user distance  &   & \\
\hline
RTCS  &  C2V        &   cellular radius    &  $\le$10Hz   &  100ms-10sec \\
\hline
CSS   & V2V,V2C,C/I2V  & cellular radius,  & event-driven,     &  1sec-1min \\
      &                & inter-vehicle gaps  & periodic at $\le$1Hz  &   \\
\hline
DID   & C2V,V2C      & cellular radius       & event-driven   &  best-effort \\
\hline
GPA   & V2V,V2C/I,C/I2V  & inter-vehicle gaps,           & event-driven,         &   100msec+, best-effort \\
      &                  & V2I distance, cellular radius & periodic at $\le$1Hz  &  \\
\hline
ESS   &   V2V,V2I,I2V   &   100m-300m    &    $\sim$10Hz   &   $\sim$100msec  \\
\hline
CDPA  &  V2C         &  cellular radius  & event-driven,          & best-effort, 5-10minutes \\
      &              &                   & periodic at $\ge$10Hz  & for powertrain fault report \\
\hline
\end{tabular}
\end{center}
\caption{Communication requirements of vehicle automation applications}  \label{table:infotainment}
\end{table}

\subsection{A multi-scale spatiotemporal perspective}

\begin{figure}[!htbp]
\centering
\includegraphics[width=\figWidth]{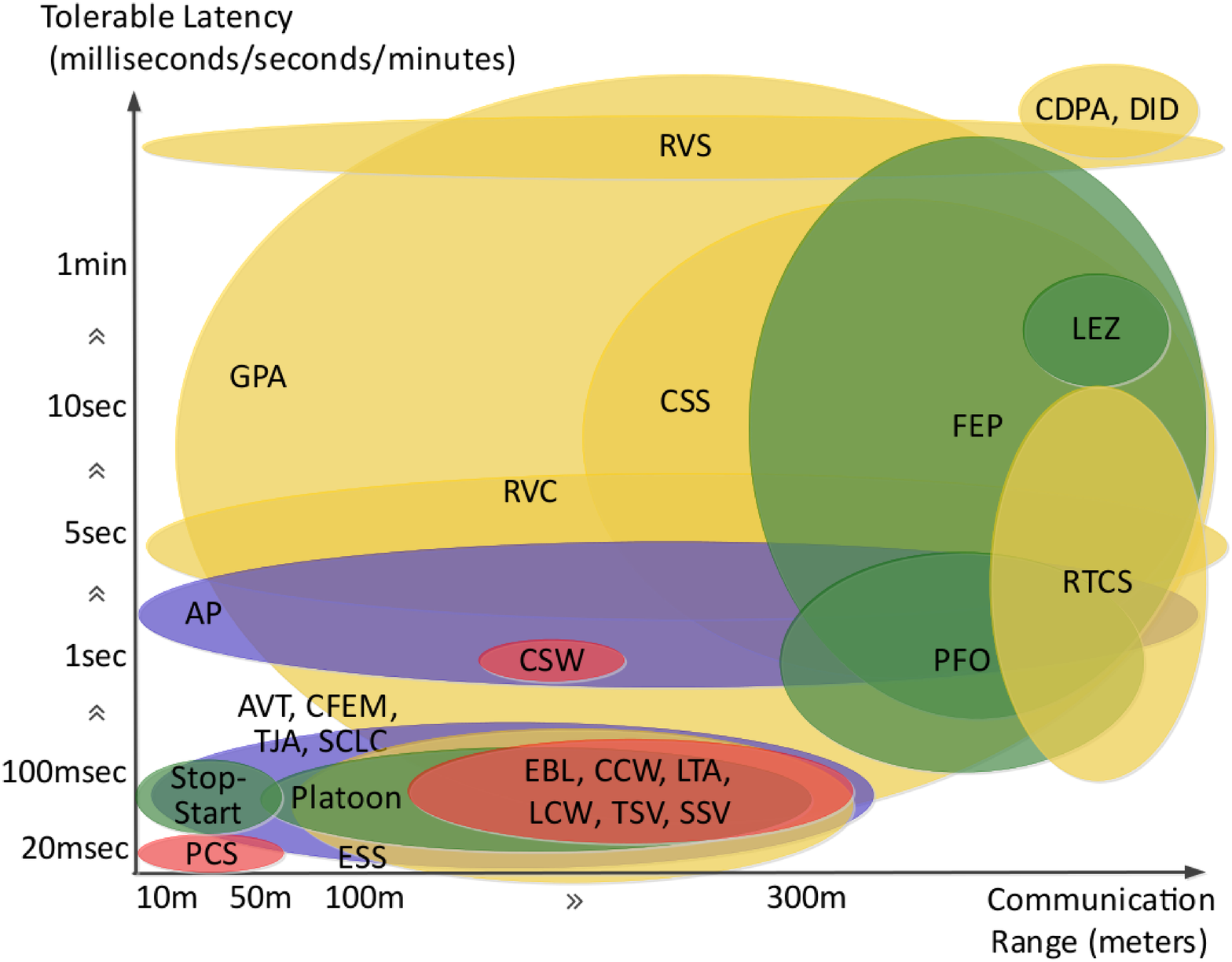}
\caption{A multi-scale spatiotemporal perspective of CAV applications} \label{fig:VNet-apps-spatiotemporal-view}
\end{figure}

From the discussions above, we observe that different CAV applications have different spatiotemporal requirements on information exchange. For reasoning about the communication requirements and strategies for enabling different CAV applications, accordingly, it is informative to organize the different application along their spatiotemporal scales. We summarize this multi-scale spatiotemporal perspective in Figure~\ref{fig:VNet-apps-spatiotemporal-view}.\footnote{The figure is not drawn to scale.}
We see that, for active safety, fuel economy and emission control, and networked vehicle automation in general, the spatial and temporal scales of communication requirements are correlated. 

\section{Analysis of vehicular wireless networking} \label{sec:networks}

Two basic paradigms of vehicular wireless communication are vehicle-to-vehicle (V2V) communication and cellular communication, and one basic question is which paradigm is more suitable for the different connected and automated vehicle (CAV) applications. 
	In what follows, we first qualitatively compare the V2V and cellular paradigms, and then we quantitatively analyze the capacity of V2V networks, based on which we make recommendations on reasoning about the paradigm-choice problem from a multi-scale spatiotemporal perspective.

\subsection{Cellular vs$.$ V2V networks}  \label{subsec:cellular-vs-v2v}

In what follows, we qualitatively analyze the strengths of cellular and V2V networks in supporting different CAV applications. Focusing on the technical aspect of the topic and with a goal of shedding light on longer-term science and technology strategy planning, our analysis is not confined by today's cellular and V2V technologies. 
	Accordingly, we model cellular networks as those wireless networks where mobile stations (e.g., vehicles) communicate with one another, the transportation infrastructures, and the Internet through cellular base stations which were mostly statically deployed, and we model V2V networks as those networks where vehicles communicate with one another and the transportation infrastructure through vehicles and the transportation infrastructures themselves without involving cellular base stations in data exchange. By this approach, the mode of cellular networks where mobile stations directly communicate with one another without involving base stations in data exchange (e.g., in the envisioned 5G cellular networks) is treated as a type of V2V network. 
	Focusing on comparative capacity analysis of cellular and V2V networks, we also assume that cellular and V2V networks have access to the same amount of wireless spectrum. 

Compared with V2V networks, cellular networks have the following advantages:
\begin{itemize}
\item With cellular base stations connected to the rest of the Internet, cellular networks can enable easy access from vehicles to the cloud resources and vice versa. Therefore, cellular networks are suitable for CAV applications that need to involve the Internet cloud resources, including cloud-assisted vehicle control and infotainment applications. 

\item Cellular base stations are usually elevated above the ground, thus the channel conditions between cellular base stations and vehicles may be better than if the cellular base stations were put on ground. With the availability of multi-hop networking and relay-based collaborative communication, however, this advantage of cellular networks may become less important over time.
\end{itemize}
There have been arguments that cellular networks do not have the challenge that V2V networks have in terms of market penetration. This is only partially true. It is the case that today's deployed cellular networks can already support many CAV applications (e.g., low emission zone and remove vehicle status applications) in the regime of large spatiotemporal scales of communication requirements. Designed for traditional consumer electronics markets, however, today's cellular networks (including LTE) were not designed for mission-critical, real-time CAV control in the regime of small spatiotemporal scales of communication requirements, and they cannot ensure predictable reliability, timeliness, and throughput as required by many CAV control applications. For mission-critical, real-time vehicle control applications that are important for transportation safety, efficiency, and comfort, therefore, cellular networks also have to face the challenge of deploying new equipment and/or capabilities.

Compared with cellular networks and considering a spatial scale at the coverage of a cellular base station, V2V networks have the following strengths in supporting CAV applications:
\begin{itemize}
\item V2V enables smaller delay in inter-vehicle communication for the following reasons: 1) V2V networks enables direct communication between vehicles without having to go through a base station as in cellular networks, thus eliminating the delay introduced by the base station relay in terms of wireless transmission and reception delay as well as software, processing delay at the base station; 2) with the same transmission power, the shorter links in direct V2V communication enable higher signal-to-noise-ratios (SNRs) at receivers and thus higher success rate in packet delivery, which in turn reduces the number of retransmissions needed to deliver a packet and thus reduces delay; 3) when transmission power is controlled according to link length, the smaller contention region in direct V2V communication reduces channel access delay. In addition to helping reduce delay, the smaller contention region in V2V communication also enables higher throughput by reducing channel contention time and improving channel spatial reuse. 

For communication between vehicles along a linear/cross roads (as envisioned for active safety and platoon applications), in particular, the number of vehicles contending for channel access is proportional to the length of roads covered by V2V or cellular communication signals. Assuming that the ratio of interference range\footnote{The range within which the concurrent transmissions by two vehicles are regarded as interfering with each other.} to communication range is similar in V2V and cellular networks, the ratio of the timeliness\footnote{Timeliness is defined as the inverse of delay.} and per-vehicle throughput in V2V networks to those in cellular networks is approximately twice the ratio of the radius of a base-station-covered cell to the communication range of direct V2V communication.

\item The transmission power and thus the contention region of V2V communication can be adaptively tuned according to the required communication range (e.g., as determined by inter-vehicle distance), thus enabling further enhancement and adaptation of communication throughput and timeliness. This flexibility of capacity adaptation is also critical for the co-design of CAV control and wireless networking. In cellular networks, however, base stations are mostly fixed, and this fact confines the flexibility of communication range adaptation and thus hampers effective capacity utilization; for instance, even if two close-by vehicles need to communicate with each other, they have to transmit at a high-enough power to reach the cellular base station. 

\item Many inter-vehicle communication required by CAV applications is broadcast in nature (e.g., for active safety and extended surrounding sensing), and this matches well with the broadcast nature of V2V wireless communication. Existing cellular architecture, however, is designed around point-to-point communication, and it introduces unnecessary overhead in supporting broadcast. 

\item V2V communication does not rely on road-side infrastructures (e.g., base stations), and, for CAV applications in the regime of small spatiotemporal scale of communication requires, V2V networks provide the communication service whenever and wherever need be, even in areas (e.g., rural areas) of little or no cellular coverage.
 
\end{itemize}

\subsection{Per-vehicle transmission capacity and delay in V2V networks}

To elaborate on the analysis of Section~\ref{subsec:cellular-vs-v2v}, here we analyze the per-vehicle transmission capacity and delay in V2V networks. 

To understand the impact of the road and traffic condition,  we consider scenarios where the number of lanes (denoted by $N$) in a road segment ranges from 2 to 8, and the inter-vehicle gap (denoted by $D$) ranges from 6 meters to 300 meters representing different degrees of traffic congestion and speed \cite{Bai:highway-Poisson-distribution}. For other network parameters, we consider the following typical settings:
\begin{itemize}
\item Transmission range: $R$ meters (default 300 meters);
\item Ratio of interference range to transmission range: $I$ (default 2 as a typical value as shown by Zhang et$.$ al \cite{PRK});
\item Wireless channel capacity: $C$ Mbps (default 27Mbps, i.e., the maximum rate of the current IEEE 802.11p-based DSRC technology); 
\item Overall capacity utilization ratio to account for protocol overhead: $U \in [0, 1]$   (default 0.9)
\item Packet length: $L$ bytes (default 400 bytes)
\end{itemize}

Given a specific value of $N$ and $D$ and considering a straight road, the number of vehicles interfering with the transmission by a vehicle is $(\frac{2RI}{D}+1)N - 1$, and the effective overall capacity shared by a vehicle and its interfering vehicles is $CU$. Thus the average per-vehicle transmission capacity $T$ (Mbps) is as follows:\footnote{For simplicity, we do not consider intersections, and we assume all the vehicles are along a single road.}
\begin{equation} \label{eqn:capacity}
T = \frac{CU}{(\frac{2RI}{D}+1)N},
\end{equation} 
And the average per-packet transmission delay $d$ (msec) is as follows:
\begin{equation} \label{eqn:delay}
d = \frac{8L}{1000T}.
\end{equation}

With the default parameter settings, we have Tables~\ref{table:capacity} and \ref{table:delay} for the average per-vehicle capacity and per-packet delay respectively.
\begin{table}[ht]
\begin{center}
\begin{tabular}{| l | c | c | c | c |}
\hline
    & \multicolumn{4}{|c|}{ Number of lanes $N$ }  \\   
\hline
Inter-vehicle gap $D$ (meters)  &  2	&   4	 &    6	   &   8   \\
\hline
6   &  0.0604   &  0.0302  &  0.0201 &   0.0151 \\
\hline
20  &  0.1992   &  0.0996  &  0.0664 &   0.0498 \\
\hline
50  &  0.4860   &  0.2430  &  0.1620 &   0.1215 \\
\hline
100 &  0.9346   &  0.4673  &  0.3115 &   0.2337 \\
\hline
200 &  1.7357   &  0.8679  &  0.5786 &   0.4339 \\
\hline
300 &  2.4300   &  1.2150  &  0.8100 &   0.6075 \\
\hline
\end{tabular}
\end{center}
\caption{Average per-vehicle transmission capacity (Mbps)}  \label{table:capacity}
\end{table}
\begin{table}[ht]
\begin{center}
\begin{tabular}{| l | c | c | c | c |}
\hline
    & \multicolumn{4}{|c|}{ Number of lanes $N$ }  \\   
\hline
Inter-vehicle gap $D$ (meters)  &  2	&   4	 &    6	   &   8   \\
\hline
6   & 52.9383  & 105.8765  & 158.8148  & 211.7531 \\
\hline
20  & 16.0658  & 32.1317   & 48.1975   & 64.2634 \\
\hline
50  & 6.5844   & 13.1687   & 19.7531   & 26.3374 \\
\hline
100 & 3.4239   & 6.8477    & 10.2716   & 13.6955 \\
\hline
200 & 1.8436   & 3.6872    & 5.5309    & 7.3745 \\
\hline
300 & 1.3169   & 2.6337    & 3.9506    & 5.2675 \\
\hline
\end{tabular}
\end{center}
\caption{Average per-packet transmission delay (msec)}  \label{table:delay}
\end{table}
We see that the per-packet delay is less than 100msec in most cases except for the very high-density settings of $D=6m$ and $N \ge 4$. To further increase per-vehicle capacity and to reduce per-packet delay, we can use the following mechanisms:
\begin{itemize}
\item It is expected that, in settings of denser vehicle distributions, the vehicle speed will be lower and the required communication range can be reduced. Therefore, we can adaptively control the transmission range (e.g., through transmission power control) to reduce the number of interfering vehicles and thus increase the per-vehicle capacity. For the case of $D=6m$ and $N=8$, for instance, using a transmission range of 60 meters (i.e., 10 times the inter-vehicle distance) instead of 300 meters will increase the per-vehicle capacity and reduce the per-packet delay by a factor of about 5. Note that, since it is difficult/impossible to adaptively change the locations of cellular base stations on the fly, it is easy to implement adaptive transmission range control in V2V networks but not in cellular networks.

\item Using multiple wireless channels instead of a single channel will also enhance capacity and reduce delay.
\end{itemize}
Another observation from this analysis is that the capacity of V2V networks is only barely enough to support today's application requirements (especially in dense vehicle traffic scenarios) even with a high capacity utilization ratio of 0.9. Contention-based channel access control such as that in the IEEE 802.11p-based DSRC usually has a much lower capacity utilization ratio (e.g., less than 0.5), thus TDMA-based channel access control \cite{PRKS} is expected to be more suitable for CAV applications at the small spatiotemporal scale of communication requirements.

\subsection{A multi-scale spatiotemporal perspective}

\begin{figure}[!htbp]
\centering
\includegraphics[width=\figWidth]{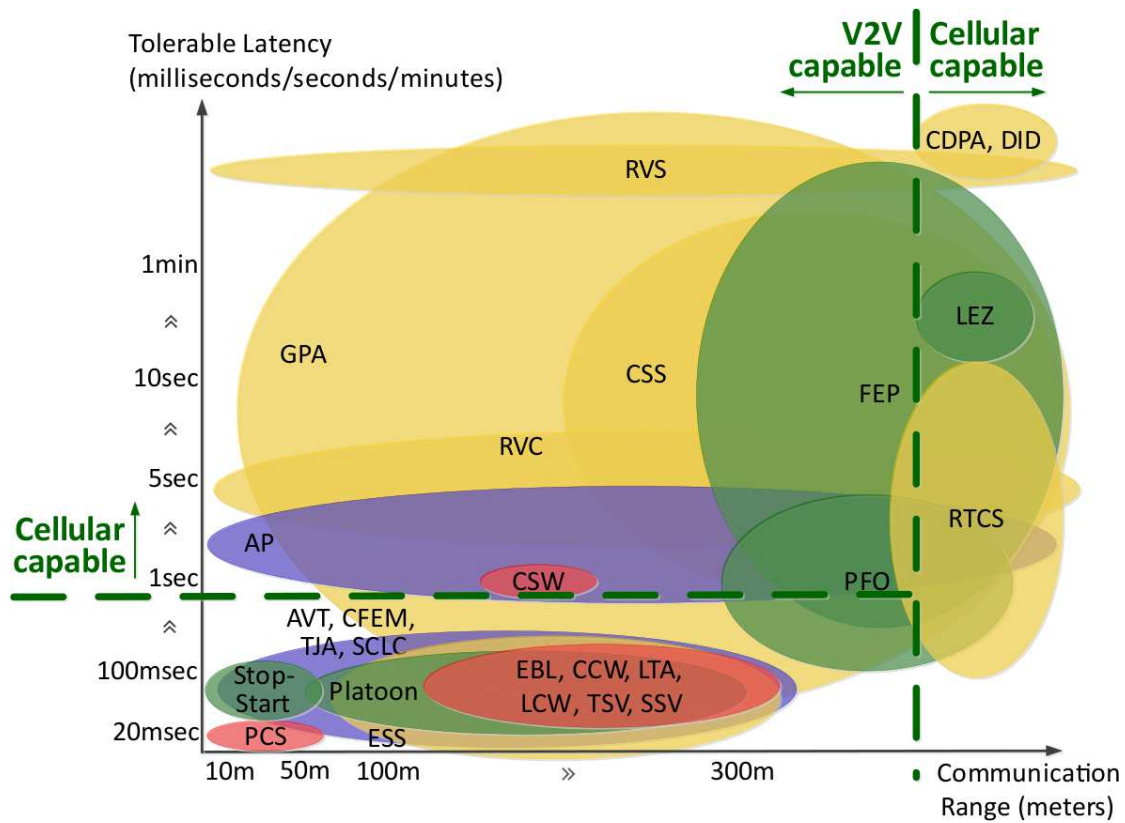}
\caption{A multi-scale spatiotemporal perspective of wireless networking for CAV applications} \label{fig:VNet-apps-networking-alternatives}	
\end{figure}

From the above analysis, we see that the direct inter-vehicle communication in V2V networks and the flexibility of adaptive transmission range and interference range control makes V2V networks suitable for those CAV applications in the regime of small spatiotemporal scales of communication requirements, and cellular networks are more suitable for those CAV applications in the regime of large spatiotemporal scales of communication requirements. We summarize this multi-scale spatiotemporal perspective in Figure~\ref{fig:VNet-apps-networking-alternatives}.\footnote{The figure is not drawn to scale.}
The fact that both V2V networks and cellular networks are suitable for those CAV applications in the regime of small spatial scales but medium-to-large temporal scales of communication requirements implies the opportunity of jointly utilizing the capacity of V2V and cellular networks for those applications.

\section{Concluding remarks} \label{sec:concludingRemarks}

We have examined the wireless communication requirements by canonical CAV applications in active safety, fuel economy and emission control, vehicle automation, and infotainment. We have observed that the spatiotemporal framework can serve as a good mechanism of reasoning about the wireless communication requirements of CAV applications. 
	Using the spatiotemporal framework, we have shown that V2V networks are effective in supporting CAV applications in the regime of small spatiotemporal scales of communication requirements, and, in the small spatiotemporal scale regime, time-division-multiple-access (TDMA) \cite{PRKS} is expected to be more suitable than contention-based channel access control such as that in IEEE 802.11p. 
	We have also observed the opportunity of jointly using the V2V and cellular networks in the small spatial scale but medium-to-large temporal scale regime. 
	These findings suggest the benefits and need for effectively utilizing V2V networks in realizing the CAV vision, incorporating TDMA as a basic channel access control mechanism in V2V networks, and leveraging the heterogeneous wireless networks available to CAV.

\section*{Acknowledgment} 

We thank K$.$ Venkatesh Prasad for hosting the sabbatical visit of Hongwei Zhang and for his insightful comments on a preliminary version of this article. We thank Qing Ai for discussing typical cellular radii in LTE networks.

{\small
\bibliographystyle{plain}
\bibliography{references}
}




\end{document}